\documentclass{aa}
\usepackage{graphicx}

\def\dg{\hbox{$^{\circ}$}}
\def\fdg{\hbox{$.\!\!^{\circ}$}}

\def\fam{\hbox{$.\mkern-4mu^\prime$}}

\newcommand{\aaps}{A\&A}
\newcommand{\aas}{A\&AS}
\newcommand{\apj}{ApJ}
\newcommand{\apjs}{ApJS}

\newcommand{\aj}{AJ}

\begin{document}
\title{
Is the Cygnus Loop two  supernova remnants?
}
\author{B. Uyan{\i}ker\inst{1}, W. Reich\inst{2},
        A. Yar\inst{1}, R. Kothes\inst{1,3}, and E. F\"urst\inst{2}}

\offprints{B. Uyan{\i}ker}

\institute{National Research Council,
           Herzberg Institute of Astrophysics,
           Dominion Radio Astrophysical Observatory,
           P.O. Box 248, Penticton, British Columbia, V2A 6K3, Canada
         \\  \email{bulent.uyaniker@.nrc.ca, aylin.yar@.nrc.ca,
                    roland.kothes@nrc.ca}
\and
           Max-Planck-Institut f\"ur Radioastronomie,
           Auf dem H\"ugel 69, 53121 Bonn, Germany
          \\   \email{wreich@mpifr-bonn.mpg.de,
                      efuerst@mpifr-bonn.mpg.de}
\and
           Department of Physics and Astronomy,
           The University of Calgary,
           2500 University Dr. NW, \\
           Calgary, AB,
           T2N 1N4 Canada
}

\titlerunning{ Is the Cygnus Loop two  supernova remnants? }
\authorrunning{Uyan{\i}ker et al.}
\date{Received 2002; accepted }

\abstract{ The  Cygnus Loop is  classified as a  middle-aged supernova
remnant  (SNR)  located   below  the  Galactic  equator  ($\ell=74\dg,
b=-8\fdg6$)  and 770~pc  away from  us.   Its large  size and  little
confusion  with Galactic  emission makes  it  an ideal  test ground  for
evolutionary  and structural  theories of  SNRs.  New  radio continuum
mapping  of the  Cygnus Loop  at 2695  MHz with  the  Effelsberg 100-m
telescope provides  indications that the  Cygnus Loop consists  of two
separate SNRs. Combining this result  with  data
from the literature we argue that  a secondary SNR exists in the south
with  a  recently detected  neutron  star  close  to its  center.  Two
interacting SNRs  seem to be the best explanation to account for the
Cygnus  Loop observations at  all wavelengths.
 }
\maketitle

\section{Introduction}

The unusual shape of the Cygnus Loop has been the focus of interest,
because this SNR, by being close to us ($\sim$770
pc, Minkowski \cite{min58}) and being relatively unaffected by the
complex Cygnus region emission, provides most detailed information on
middle-aged SNRs.  Having these properties, the Cygnus Loop is in
general considered to be the laboratory of SNR studies in the X-ray,
optical, infrared, and radio.

New polarimetric data towards the Cygnus Loop at 2695 MHz ($\lambda$=11
cm) reveal that this remnant in fact consists of two separate
SNRs. The analysis of these data together with available
information on the rotation measure (RM), X-ray, optical, and
infrared observations of the Cygnus Loop are used to constrain
the characteristics and the relation of both SNRs.

\section{Radio data at 2695 MHz}

We refer to a new sensitive radio map obtained with the Effelsberg 100-m
telescope at 2695 MHz with an angular resolution of $4\fam3$ including
linear polarization (Fig.~\ref{fig1}).  The map was observed as part
of a multi-frequency continuum and polarization study (Uyan{\i}ker et al.,
in prep.).  The shape of the
Cygnus Loop is circular in the north with an extension to the south.
The southern part of the remnant resembles a crescent whose
western part is incomplete and is mixing with weak peripheral
emission.  Nevertheless, the southern crescent is brighter than most
of the northern portions of the Cygnus Loop.  The difference in total
intensity between these two regions and
spectral index variations of $\Delta\alpha\sim0.2$ led Green (\cite{gre90})
to conclude that the northern part of the remnant is governed by a
compression of the magnetic field whereas the southern part results most
likely from shock acceleration. The morphological appearance of the
total intensity emission in the north also supports the idea that the
shock of the explosion compresses and deforms the magnetic field and
thus the remnant carries the signature of the local magnetic field
(Green \cite{gre84}). In this way the partial shells of the remnant
are aligned along the direction of the magnetic field.

The total intensity and polarization images (Fig.~\ref{fig1}) reveal a
new feature of the Cygnus Loop.  There is a diffuse radio plateau
enveloping the conventional bright radio appearance of the
remnant. This plateau is the counterpart of those in the
X-ray and infrared emission. It is prominently seen towards the
north and north-eastern shell of the remnant. Thus with the detection
of the radio plateau a general agreement in the radio, X-ray and
infrared appearance of the remnant for its northern shell is reconciled.
This plateau was not detected in very early Effelsberg observations
(Keen et al. \cite{kee73}).

\begin{figure*}\centering

\vspace{8cm}
\caption{Radio images of the Cygnus Loop at 2695 MHz at a
resolution of $4\fam3$. Both images are in the equatorial coordinate
system (J2000). The total intensity map is shown on the left panel,
where grayscale coding represents continuous intensity levels from 0
mK T$_{\rm B}$ (white) to 500 mK T$_{\rm B}$ (black).  The overlaid
bars are electric field vectors proportional to the polarized
intensity. The polarized intensity image of the same region is displayed
on the right hand panel. Here the grayscale coding represents levels
from 0 mK T$_{\rm B}$ (white) to 90 mK T$_{\rm B}$ (black).
\label{fig1}}
\end{figure*}

\section{Polarization data}
The new polarization image of the remnant shows a fair amount of
polarization towards the south (up to 30\%), as well as the prominent
filament almost dividing the northern shell into two symmetrical half
shells. However, there is a clear distinction in the distribution of
the polarized emission. The percentage polarization in the northern part is
drastically lower than in the south, especially away from the
polarized central filament in the north. RM calculations from the four bands
of the 1420~MHz Dominion Radio Astrophysical Observatory (DRAO)
archival interferometric maps together with the presented 2695 MHz
data (Uyan{\i}ker et al., in prep.), as well as a previous
analysis by Kundu \& Becker (\cite{kun72}), suggest that RMs vary
across the remnant between $-10$ to $-36$ rad/m$^2$. The mean
is about $-20$ rad/m$^2$. Such a RM will rotate the observed polarization
angle at 2695~MHz by about $15\dg$.  Therefore the 2695~MHz map
traces the orientation of the magnetic field towards the remnant
closely.

Consequently, the asymmetric polarization distribution between the
southern and the northern part is not due to a RM gradient, but rather
intrinsic. Differences in magnetic field configuration
between south and north indicate the action of different mechanisms in
these regions. Furthermore, while the polarized intensity in the
southern part delineates a secondary circular shell corresponding to
the  previously called 'break-out' region with magnetic field vectors aligned
tangentially, the northern part displays a rather complicated
appearance.  We emphasize that the upper part of the polarized
southern shell overlaps with the northern shell, but results from a shock
originating in the south. This contradicts the previous interpretation that
a shock from the northern area breaks out into the southern region.
Indeed, Tenorio-Tagle et al. (\cite{ten85}) proposed that the
northern shell of the Cygnus Loop is the breakout of a SNR in the
south. However, their model has difficulties to explain the high
X-ray emission from the northern shell. Furthermore, this model needs
ambient material of higher
density in the south -- which is neither observed in
the infrared (Arendt et al. \cite{are92}), in \ion{H}{i} (DeNoyer
\cite{den75}) or CO emission (Scoville et al. \cite{sco77}).

\begin{figure*}\centering
\vspace{8cm}
\caption{Left panel: Sketch of the Cygnus Loop  showing radio emission (thin
lines) and 0.25~keV X-ray emission from ROSAT data (thick lines,
convolved with a Gaussian beam of $3^{\prime}$).  The positions and extents
of the two SNRs are represented by dashed ellipses.  The magnetic
field vectors, obtained by using the polarization image given in
Fig.~\ref{fig1} and corrected for Faraday rotation are overlaid as
bars. Note that pixels with low signal to noise ratio are omitted and
the length of the vectors are proportional to RM.
The vectors of strong point sources are screened to avoid
confusion.  The direction of the large-scale magnetic field in this
area of the Milky Way is shown by a dashed straight line, which passes
through the polarized central filament in the north.  The neutron star
in the southern part is indicated by a plus sign.  The two prominent
optical filaments, NGC~6992 and NGC~6960 are also labeled.  Right
panel: X-ray emission from the Cygnus Loop at
0.25~keV convolved with a $3^{\prime}$ beam. The X-ray intensities
range from from 0.002~counts s$^{-1}$ arcmin$^{-2}$ to 0.25~counts
s$^{-1}$ arcmin$^{-2}$.  Overlaid thick contours represent polarized
intensity levels starting at 40 mK T$_{\rm B}$ running in steps of 40 mK
T$_{\rm B}$.
\label{fig2} }
\end{figure*}

We conclude from the above arguments that neither a southern nor a
northern breakout explains the available observations. There is
strong evidence for two separate remnants rather than a single
object.

The presence of the anomalous X-ray source AX J2049.6+2939 detected by
the ASCA satellite (Miyata et al. \cite{miy01}) at the center of the
southern remnant provides further proof that this part is
independent from the large northern shell. The X-ray spectrum of this
source results in a photon index of $-2.1\pm$0.1.  Although no pulsed X-ray
or radio emission was detected towards this object, Miyata et al.
(\cite{miy01}) conclude that this object is a stellar remnant,
probably a neutron star.  Any relationship between this X-ray source
and the southern part strengthens the idea that the Cygnus Loop
consists of two SNRs.

\section{Optical, X-ray and  infrared  data}

Optical images (Levenson et al. \cite{lev98}) reveal differences
between the northern and the southern part of the Cygnus Loop. For
example, H$\alpha$ and \ion{O}{iii} regions are more extended in the north than
in the south as a consequence of differences in the interaction with the
surrounding gas.

The X-ray (Ku et al. \cite{ku84}, Levenson et al. \cite{lev98}) and
infrared emission (Braun \& Strom \cite{bra86}) of the northern part
is consistent with a limb-brightened shell of hot gas, while the southern
source is almost invisible. This difference was explained as a sign for
different mechanisms acting in the north and the south of the Cygnus Loop
(Green \cite{gre90}).

We notice that at all wavelengths there are different emission
characteristics in the northern part compared to the southern
region.

Aschenbach \& Leahy (\cite{asc99}) have interpreted the southern part
of the remnant as a breakout into a lower density medium. Their main
reason was the smooth change of the X-ray emission from the northern to the
southern area. However, a smooth transition may also result from two
SNRs seen in superposition. In
Fig.~\ref{fig2} (right panel) we display the soft X-ray image with
overlaid polarization intensity contours showing a relation
of minimum X-ray emission and maximum polarized emission in the
area where the southern shell superimposes with the
northern one. In addition there are two regions from $\rm \sim \alpha, \delta =
20^{h} 54^{m}, 30^{\circ}$ to $\rm \sim \alpha, \delta =
20^{h} 53^{m}, 29^{\circ} 30'$ at the eastern and at
$\rm \sim \alpha, \delta = 20^{h} 48^{m}, 29^{\circ} 45'$ at the
western side, where X-rays are enhanced and also the periphery
of both shells superimpose. These findings strongly suggest that
both SNR shells are not just seen in superposition, but are in
physical interaction. Enhanced soft X-rays might emerge from the
deceleration of the shock waves due to the collision of the SNRs.
On the other hand it is difficult to understand the minimum X-ray
emission in the overlapping region of both SNRs. We note in that context
that Williams et al. (\cite{wil97}), who studied the colliding SNRs
DEM L316 in the Large Magellanic
Cloud, did not find X-ray emission in the region of overlap of
both SNRs. We conclude that details of the conditions in the interacting
regions are not well constrained and need a more detailed investigation

According to the interaction scenario the SNR in the north exploded first
and created a complete circular, shell-type SNR. A second SN event
took place at the position of AX J2049.6+2939, not far away from the
already expanding shell of the first explosion. Shortly after the
second explosion the shock waves of the SNRs collided. The younger
southern SNR had a higher energy density than the northern one,
swept-up the shell of the first remnant and
bent it inward.  This way the pronounced upper polarized shell of the
southern SNR was created. The collision of the two shells also
created the X-ray extensions encompassing the polarized emission (see
Fig.~\ref{fig2}).  Comparable distances of the SNRs, however, cannot
be directly utilized to estimate the quantitative ages of these SNRs
by simply comparing their sizes, since at least the northern remnant
is suspected to be expanding in a pre-existing cavity (Charles et
al. \cite{cha85}, Levenson et al. \cite{lev97}).

\section{Discussion}
We have presented new radio continuum measurements of the Cygnus Loop
including linear polarization and analyzed the differences between its
southern and northern part. These differences are
rather obvious by inspecting the polarized intensity image.  We also
report on the detection of a weak radio plateau positionally coinciding
with those observed in the infrared and in X-rays. The following
reasons led us to conclude that two SNRs make up the Cygnus Loop and
are in interaction.

\begin{list}{00}{}
\item[1)] The radio morphology of the Cygnus Loop differs in the south
and north. There is an elliptical shell in the north, including a
radio plateau corresponding to X-ray and infrared emission.  A bright
southern extension is present.

\item[2)]  The distribution of polarization intensity differs
completely in the south and north. RMs are not high enough to
depolarize the polarized emission at 2695~MHz significantly,
indicating an intrinsic difference in the magnetic field
configuration.

\item[3)]  The tangential magnetic field structure indicates a regular
magnetic field configuration in the southern part, but disturbed in the
region of overlap, whereas it is more
chaotic in the north.

\item[4)] The radio spectral index changes across the Cygnus Loop
indicating a steeper spectrum (less compression) for the southern part,
which in turn implies a different acceleration mechanism.

\item[5)] The northern part has different characteristics of the optical
filaments compared with the southern region.

\item[6)] There is weak or no X-ray and infrared emission in the south
in contrast to the strong radio appearance.

\item[7)] X-ray enhancements seen at the superposition of both
shell's periphery and decreased X-ray emission in the overlap regions
clearly indicate an interaction of both SNRs.

\item[8)] Finally, there is evidence for the existence of a stellar
remnant, probably a neutron star, almost exactly at the center of the
southern shell.

\end{list}

The northern remnant is centered at
$\sim(\alpha,\delta)=(20^{\rm h}~51\fm36$, $31\degr~3\arcmin$) and
its extent is about $3\fdg0\times2\fdg6$. The southern remnant, whose
extent is about $1\fdg4\times1\fdg8$, is centered at
$\sim(\alpha,\delta)=(20^{\rm h}~49\fm56$, $29\degr~33\arcmin$).
Thus we designate these remnants as G74.3$-$8.4 and G72.9$-$9.0
according
to their Galactic coordinates, respectively.  The results from a
comparative investigation of new radio data at other frequencies
including a spectral index analysis and a detailed description of the
polarization characteristics will be given elsewhere.

\begin{acknowledgements}
We thank Nancy Levenson for providing us with the 0.25 keV ROSAT
mosaic and Tom Landecker for critical reading of the manuscript and
useful discussions.
\end{acknowledgements}

\end{document}